\documentclass[aps,prl,twocolumn,superscriptaddress,showpacs,letter]{revtex4}
\usepackage{amsfonts,amssymb,amsmath}
\usepackage{color}
\usepackage{graphicx}
\usepackage{dcolumn}
\usepackage{bm}

\begin{document}
\title{Exact results of the quantum phase transition for the topological order}

\author{Jing Yu}
  \affiliation{Department of Physics, Beijing Normal University, Beijing 100875, China}

\author{Su-Peng Kou}
  \email{spkou@bnu.edu.cn}
  \affiliation{Department of Physics, Beijing Normal University, Beijing 100875, China}

\author{Xiao-Gang Wen}
  \affiliation{Department of Physics, Massachusetts Institute of Technology, Cambridge, Massachusetts 02139}

\begin{abstract}
In this paper{\em \ }a duality between the $d=2$\ Wen-plaquette
model in a transverse field and the $d=1$\ Ising model in a
transverse field{\em \ }is used to learn the nature of the quantum
phase transition (QPT) between a spin-polarized phase and a
topological ordered state with the string-net condensation. The QPT
is not induced by spontaneous symmetry breaking and there are no
conventional Landau-type local order parameters. Instead the
string-like non-local order parameters are introduced to describe
the QPT. In particular, the duality character between the
open-string and closed string for the QPT are explored.


\end{abstract}

 \maketitle

Landau developed a sysmatical theory for classical statistical
systems. Different orders are characterized by different symmetries.
The phase transition between different phases are always accompanied
with symmetry breaking. However, Landau's theory cannot describe all
the continuous phase transition such as the transitions between
quantum ordered phase\cite{wen}. Topological ordered state is a
special quantum order with full gapped excitations. The first
nontrivial example of topological order is the fractional quantum
hall (FQH) states which cannot be described by broken symmetries
\cite{wen}. Recently, people pay attention to the quantum phase
transition (QPT) for the topological orders. It is known that there
exists a quantum phase transition between a spin-polarized phase and
a topological ordered state. The two phases cannot be characterized
by a local order parameter. Thus such a quantum phase transition
cannot be described by the Landau theory obviously. To find a new
approach to learn the nature of such QPT becomes an interesting
issue. The quantum phase transition for the
Kitaev toric-code model in a transverse magnetic field was studied in Ref.%
 \cite{nayak}. The numerical simulations demonstrate the condensation of
`magnetic' excitations and the confinement of `electric' charges of
the phase transition out of the topological phase. And then people
find that the ``topological entropy'' $S_{\mathrm{top}}$\
\cite{Kitaev:06Levin:06,levin} serves as an order
parameter\cite{lidar,chamon}. At the QPT the ``topological entropy''
jumps abruptly from a finite number to zero out of the topological
phase. In addition, in Ref. \cite{xiang} it is revealed that the
quantum phase transitions between different topological orders can
be exactly characterized by a topological order parameter and shows
the one dimensional (1D) properties.

In this paper we focus on the Wen-plaquette model on a square lattice\cite
{wen,wen1}. The ground states of the Wen-plaquette model are $Z_2$
topological ordered state which is similar to that for the kitaev toric-code
model \cite{kitaev}. After adding the transverse field, a quantum phase
transition occurs by raising the strength of the transverse field. However,
the model cannot be solved exactly. Fortunately, we find\emph{\ }a duality
between the $d=2$\ Wen-plaquette model in a transverse field ( transverse
Wen-plaquette model) and the $d=1$\ Ising model in a transverse field (
transverse\ Ising model). These results indicate a new type of QPT between a
topological ordered phase and a non-topological ordered phase, including the
non-local order parameters, the 1D properties of the QPT, in particular,
\emph{the duality between open-string and closed-string}.

The Hamiltonian for the $d=2$ transverse Wen-plaquette model on square
lattice\cite{wen,wen1} is given as:
\begin{equation}
\mathcal{H}_{\mathrm{wen}}=g\sum_iF_i+h\sum_i\sigma _i^x  \label{wen}
\end{equation}
with $F_i=\sigma _i^x\sigma _{i+\hat{e}_x}^y\sigma _{i+\hat{e}_x+\hat{e}%
_y}^x\sigma _{i+\hat{e}_y}^y$ and $g<0.$ $\sigma _i^x,$ $\sigma _i^y$ are
Pauli matrices on sites, $i.$

For the first part of the model, it is an exact solved model\cite{wen1}. The
ground state of it is a topological order described by $Z_2$ projective
symmetry groups. And there exists three types of quasiparticles: $Z_2$
charge, $Z_2$ vortex, and fermions, of which three types of string operators
T1, T2, T3, are defined, respectively\cite{wen2}. The fermions can be
regarded as bound states of a $Z_2$ charge on even plaquette and a $Z_2$
vortex on odd plaquette. The string operator for $Z_2$ charge has a form $%
W_c(C)=\prod\limits_{m\in C}\sigma _{i_m}^{l_m}$, where $C$ is a string
connecting the even-plaquettes of the neighboring links, and $i_m$ are sites
on the string. For $Z_2$ vortex, string operator is $W_V(C)=\prod\limits_{m%
\in \tilde{C}}\sigma _{i_m}^{l_m}$ with $\tilde{C}$ as a string connecting
the even-plaquettes of the neighboring links. For an open T1 (T2) string,
the ends points represent the $Z_2$ charge ( $Z_2$ vortex). It is known that
the ground state of the topological order has a condensation of the
closed-strings as $\left\langle W_{c,\tilde{c}}(C)\right\rangle \neq 0.$
\cite{wen2} On the contrary, the open strings have mass gap without
condensation.

It is known that the Wen-plaquette model ($h=0)$ can be mapped onto nearest
neighbor Ising chains\cite{ort,hu}. This is because that the Wen-plaquette
model ( $h=0$) has an energy spectrum which is identical to that of the one
dimensional Ising chain\cite{ort}. Furthermore, we find that the transverse
Wen-plaquette model on a square lattice in Eq.(\ref{wen}) corresponds to the
one dimensional Ising chain in transverse field. In the following part of
this paper, we will show the duality and use it to learn the nature of the
QPT between a spin-polarized phase and a topological ordered state.

To obtain the duality we define $\sigma _i^x\sigma _{i+\hat{e}_x}^y\sigma
_{i+\hat{e}_x+\hat{e}_y}^x\sigma _{i+\hat{e}_y}^y=A_i,$ $\sigma _i^x=B_i.$
The two terms here have the commutation relations as: $\left[ A_i,B_j\right]
=2A_iB_j(\delta _{j,i+\hat{e}_x}+\delta _{j,i+\hat{e}_y})$ and $\left[
A_i,A_j\right] =0,$ $\left[ B_i,B_j\right] =0.$ Then we denote $A_i$ by $%
\tau _{i+\frac 12}^x$. To realize the Pauli algebra for $A_i$ and $B_j$, $%
B_j $ is denoted by $\tau _{j-\frac 12}^z\tau _{j+\frac 12}^z$. Here $\tau
_{j+\frac 12}^x,$ $\tau _{j+\frac 12}^z$ are Pauli matrices on sites, $%
j+\frac 12.$ Then the mapping between the two models is given as
\begin{equation}
\sigma _i^x\sigma _{i+\hat{e}_x}^y\sigma _{i+\hat{e}_x+\hat{e}_y}^x\sigma
_{i+\hat{e}_y}^y\longmapsto \tau _{i+\frac 12}^x,\text{ }\sigma
_i^x\longmapsto \tau _{i-\frac 12}^z\tau _{i+\frac 12}^z.  \nonumber
\end{equation}

Then we can explicitly denote the original model by the following
Hamiltonian, describing the Ising model in a transverse field in $d=1$:
\begin{equation}
\mathcal{H}_{\mathrm{wen}}\mapsto \mathcal{H}_I=-h\sum_a\sum_i\left( g_I\tau
_{a,i+\frac 12}^x+\tau _{a,i-\frac 12}^z\tau _{a,i+\frac 12}^z\right)
\label{hamising}
\end{equation}
where $a$ is the chain-index which is different for the $d=2$ transverse
Wen-plaquette model on different lattices with different boundary conditions
(see Fig.1). $h$ is an overall energy scale and $g_I=\frac 1{h_I}=\frac gh>0$
is a dimensionless coupling constant\cite{commet}. This is a 1D Ising model
along diagonal directions. By the mapping above, we find $g$-term in wen's
model corresponds to external field term in Ising model, while $h$-term
corresponds to Ising-term.

\begin{figure}
\includegraphics[width=0.48\textwidth]{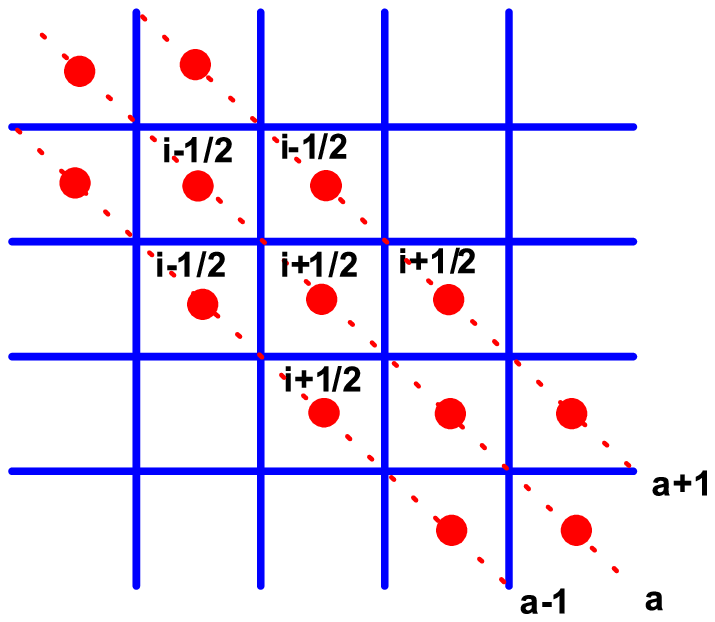}
\caption{The dotted lines are the dual Ising chains with $a$ as the
chain-index and $i$ as the site-index. }
\label{Fig.1}
\end{figure}

In addition, an important feature about the duality is the boundary
condition. The $d=2$\ transverse Wen-plaquette model on $N\times N$\ square
lattice with the open boundary condition is dual to $2N-1$\ decoupled Ising
chains ( in a transverse field). Here the chain-index $a$ is from $1$ to $%
M+N-1$. However, on $N\times N$\ square lattice with the periodic boundary
condition, the $d=2$\ transverse Wen-plaquette model $1$\ corresponds to $N$%
\ decoupled Ising chains.

\textit{Global phase diagram --- }From the duality, one can obtain the
global phase diagram for the original model by studying the transverse Ising
chain. The model in Eq.(\ref{wen}) has the same phase diagram to that for
the $d=1$ transverse Ising model described by Eq.(\ref{hamising}). Then the
original model in Eq.(\ref{wen}) has a $T=0$ quantum phase transition at $%
g_I=\frac gh=1$ from topological ordered state with ($g_I>1$), to a gapped
spin-polarized state with ($g_I<1$). For large $g_I$, $g_I>1$, the ground
state is topological order which is dual to the spin-polarized state in the
1D Ising model with $\left\langle \tau _{i+\frac 12}^z\right\rangle =0$ . On
the other hand, for small $g_I$, $g_I<1,$ the spin-polarized state is in a
superposition of $\sigma ^x$ eigenstates, with different sites uncorrected.
This state corresponds to the spin ordered state in the dual 1D Ising model
with a spontaneous magnetization $\left\langle \tau _{i+\frac
12}^z\right\rangle \neq 0$.

\label{sec:ising}\textit{The scaling law near the QPT ---} To obtain the
scaling law near the QPT, the above spin-$\frac 12$ transverse Ising model
in Eq.(\ref{hamising}) can be described by the following Hamiltonian
\begin{eqnarray}
\mathcal{H}_I &=&h\sum\limits_{a,j}[-(c_{a,j}^{\dagger
}-c_{a,j})(c_{a,j+1}^{\dagger }+c_{a,j+1}) \\
&&+g_I(c_{a,j}^{\dagger }-c_{a,j})(c_{a,j}^{\dagger }+c_{a,j})]  \nonumber
\end{eqnarray}
after employing Jordan-Wigner transformation of the spin operators to
spinless fermions\cite{lsm,mccoy}, $\tau _{a,j+\frac 12}^x=2c_{a,j}^{\dagger
}c_{a,j}-1,$ $\tau _{a,j+\frac 12}^z=(-1)^{j-1}\exp (\pm i\pi
\sum\limits_{n=1}^{j-1}c_{a,n}^{\dagger }c_{a,n})(c_{a,j}^{\dagger
}+c_{a,j}).$ In the fermionic representation for the transverse Ising chain,
the energy spectrum for fermions $E_k=\pm 2h\sqrt{(g_I-\cos (k\tilde{a}%
))^2+\sin ^2(k\tilde{a})}$ has an energy gap $E_{k=0}=2h|g_I-1|$ which tends
to zero at $g_I\rightarrow 1$. So the energy gap for the fermions must tend
to zero at the transition $m=\frac 1{2h\tilde{a}^2}(g_I-1)~\mbox{and}~~c=2h%
\tilde{a},$ where $\tilde{a}$ is the lattice spacing. Here $m$ is the mass
for the Majorana fermion which changes sign by tuning the mode across the
transition; we have chosen $m>0$ to correspond to the topological ordered
side.\cite{ising}.

\textit{The 1D properties for the QPT} \textit{--- }People have guessed that
the energy gap for the elementary excitations will always close at the QPT
for topological orders. Our results confirm the conjecture. At the critical
point, $g_I=1,$ the energy for the fermion excitation becomes $E_k=sim
c\cdot k$ where the velocity is $2h\tilde{a}$. Since the result is obtained
from the dual Ising chain in Eq.(\ref{hamising}), the energy gap between the
ground state and the first excited state is $\delta E_k\sim c\cdot \delta
k=c\cdot \frac{2\pi }L$ where $L$ is the length of the Ising chain. The
energy gap turns into zero at QPT in the thermodynamic limit. In Ref.(\cite
{lidar}) in the thermodynamic limit, it is predicted that the gap between
the ground state and the first excited state closes scales as at the
critical point $\delta E(\mathcal{N})=E_1-E_0\sim \mathcal{N}^{-\frac 12},$ $%
(\mathcal{N}=N^2).$ On a $N\times N$ lattice with periodic boundary
condition, the original model is dual to $N $ decoupled Ising chains, then
the gap is determined by the length of each Ising chain which indeed scales
as $\delta E\sim N^{-1}\sim \mathcal{N}^{-\frac 12}.$ The Fig.2 shows the
energy gap $\delta E_k$ at the critical point on different lattices from
above results and those from the exact diagonal numerical results. From the
results in Fig.2, the energy gap doesn't always scale as $\delta E(\mathcal{N%
})\sim \mathcal{N}^{-\frac 12}$\emph{. }

\begin{figure}
\includegraphics[width=0.5\textwidth]{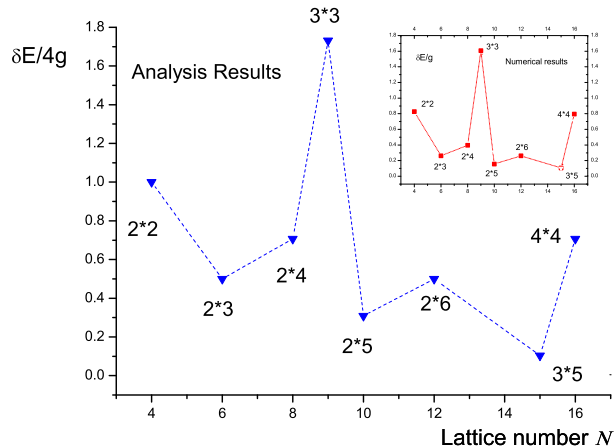}
\caption{The energy gap $\delta E$ (scaled by $4g$ ) between the ground
state and the first excited state at the critical point $g=h$ on different
lattices from analysis results. The inset shows the energy gap $\delta E$
(scaled by $g$ ) at the critical point $g=h$ on different lattices from the
exact diagonal numerical results. Here $N*M$ denotes a $N\times M$ lattice.}
\label{Fig.2}
\end{figure}

In addition, the correlation function at the QPT shows the 1D characters%
\textit{. }The long-distance limit of the static correlator for $F_iF_j$ of
the 2D transverse Wen-plaquette model can be described by the dual
correlator for $\tau _{a,i}^x\tau _{a^{\prime },j^{\prime }}^x$ of the 1D
Ising model in the transverse field. For a diagonal case, $F_iF_j=F_iF_{i-n%
\hat{e}_x+n\hat{e}_y}$, the correlation function at the QPT is given as\cite
{mccoy}\cite{ising}
\begin{equation}
\left\langle F_iF_{i-n\hat{e}_x+n\hat{e}_y}\right\rangle -\frac 2{\pi
^2}=\frac 4{\pi ^2(4n^2-1)}.
\end{equation}
Here $m$, $n$ are integer numbers.

\textit{The non-local order parameters ---} Firstly we try to use the
expectation value for $\sigma _i^x$ and $F_i$ to be the order parameters to
describe the QPT. However, $\langle \sigma _i^x\rangle $ and $\langle
F_i\rangle $ are finite in both phases but discontinue at the QPT. So one
cannot use the expectation values of $\sigma _i^x$ and $F_i$ to be the order
parameters to describe the QPT.

\begin{figure}
\includegraphics[clip,width=0.4\textwidth]{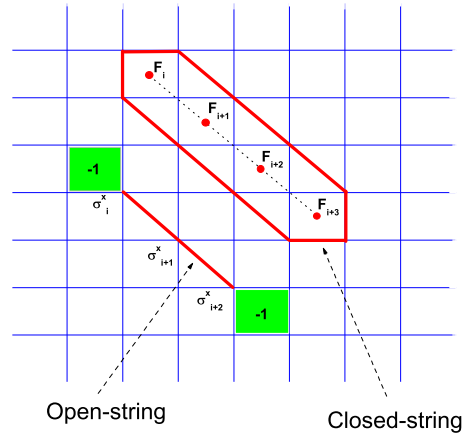}
\caption{The relationship between an open-string and a closed-string. The
hexagon black line denotes a closed-string for $F_iF_{i+1}F_{i+2}F_{i+3}.$
The black line below it is a open-string for $\sigma _i^x\sigma
_{i+1}^x\sigma _{i+2}^x,$ of which two Z2 charges (or Z2 vortexes) live at
the end denoted by the black squares.}
\label{Fig.3}
\end{figure}

Instead, we define two kinds of non-local order parameters : one is $\phi
_1=\langle \prod\limits_{i_n=1}^i\sigma _{i_n}^x\rangle $; the other is $%
\phi _2=\left\langle \prod\limits_{_{i_n=1}}^iF_{i_n}\right\rangle $. Here $%
i_n=i-n\hat{e}_x+n\hat{e}_y$ represents the sites along diagonal direction.
Let's explain the physics properties of them. On the one hand, $%
\prod\limits_{i_n=1}^i\sigma _{i_n}^x$ is a T1 or T2 open-string operator
with two ends, $i_n=1$ and $i_n=i$. On the other hand, $\prod%
\limits_{_{i_n=1}}^iF_{i_n}$ is a T1 or T2 close-string operator around the
points $i_n=1$ and $i_n=i$. Fig.3 shows the relationship between an
open-string and a closed-string. The expectation values of (open or close)
string operators\textit{\ }$\prod\limits_{i_n=1}^i\sigma _{i_n}^x$ and $%
\prod\limits_{_{i_n=1}}^iF_{i_n}$ are given by the Fig.4. From it one can
see that the QPT can be characterized in terms of the two kinds of non-local
order parameters.

\textit{The open-string - closed-string duality ---}\emph{\ }Another
important feature for the QPT between the topological order and the
non-topological order is \emph{the open-string - closed-string duality.}
From the duality relationship, $\sigma _i^x\sigma _{i+\hat{e}_x}^y\sigma _{i+%
\hat{e}_x+\hat{e}_y}^x\sigma _{i+\hat{e}_y}^y\longmapsto \tau _{i+\frac
12}^x $ and $\sigma _i^x\longmapsto \tau _{i-\frac 12}^z\tau _{i+\frac 12}^z,
$ one can define the first order parameter as $\phi _1=\left\langle \tau
_{a,1+\frac 12}^z\tau _{a,i+\frac 12}^z\right\rangle $ and the second as $%
\phi _2=\left\langle \prod\limits_{_j=i}^i\tau _{a,i+\frac
12}^x\right\rangle $ in the dual model. And $\phi _1$ is just the local
order parameter in dual model which is dual to $\phi _2$, a `disorder order
parameter' for the transverse Ising chain. This property leads to the
duality between open-string and closed string.

From Fig.4, one can see that in the topological ordered phase, $g_I>1$, we
have $\left\langle \phi _1\right\rangle =0$ and $\left\langle \phi
_2\right\rangle \sim (1-\frac hg)^{\frac 18}\neq 0.$ This result shows that
in this phase the T1 and T2 close-strings are both condensed while the
open-strings of them are not. It is consistent to the fact that the ground
state of the topological order has a condensation of the closed-strings \cite
{wen2} $\left\langle W_{c,\tilde{c}}(C)\right\rangle \neq 0$.

\begin{figure}
\includegraphics[clip,width=0.4\textwidth]{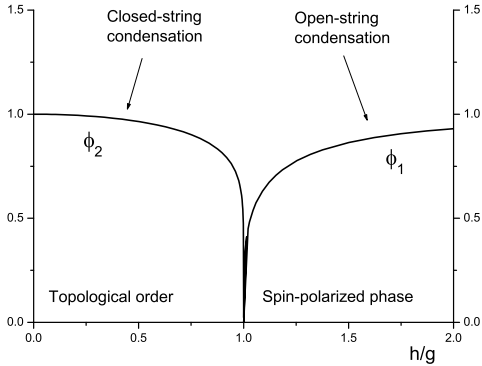}
\caption{The expectation values of (open or close) string operators\textit{\
}$\prod\limits_{i_n=1}^i\sigma _{i_n}^x$ and $\prod%
\limits_{_{i_n=1}}^iF_{i_n}$. }
\label{Fig.4}
\end{figure}

On the other hand, in the spin-polarized phase, $g_I<1$, a long range order
for dual spin correlation function of $\tau _{a,i-\frac 12}^z$ indicates the
condensation of the open-string operator near the QPT\cite{xiang}
\begin{equation}
\left\langle \phi _1\right\rangle =\lim_{i\rightarrow \infty }\langle
\prod\limits_{i_n=1}^i\sigma _{i_n}^x\rangle =\lim_{i\rightarrow \infty
}\langle \tau _{a,\frac 12}^z\tau _{a,i-\frac 12}^z\rangle \sim (1-\frac
gh)^{\frac 14}.
\end{equation}
It is known that the open-string operator, $\prod\limits_{i_n=1}^i\sigma
_{i_n}^x,$ will generate two Z2 vortexes (or Z2 charges) at both ends of the
string. The nonzero expectation value of it means that it is \emph{the
`condensation' of both Z2 vortex and Z2 charge} that breaks down the
topological order. The disappearance of the condensation of closed-string $%
\left\langle \phi _2\right\rangle =0$ or $\left\langle W_{c,\tilde{c}%
}(C)\right\rangle =0$ means that this phase is not a topological order any
more.

The situation here is much difference from that in Ref.\cite{nayak}, of
which the topological order is broken down by the condensation of `magnetic'
excitations (Z2 vortex ), and the confinement of `electric' charges (Z2
charge). Thus our results indicate there may exist different roads to break
down the topological order : the condensation for both Z2 vortex and Z2
charge or the condensation of Z2 vortex (`magnetic' excitation) together
with the confinement of Z2 charge (`electric' charge).

In summary, a duality between the $d=2$\ Wen-plaquette model in a transverse
field ( transverse Wen-plaquette model) and the $d=1$\ Ising model in a
transverse field ( transverse\ Ising model) are uncovered. By the duality,
we find that the $d=2$ transverse Wen-plaquette model on square lattice
undergoes a continuous phase transition. The quantum phase transition
between a spin-polarized phase and a topological ordered state with the
string-net condensation is not induced by spontaneous symmetry breaking and
there are no conventional Landau-type local order parameters. Instead, the
QPT can be characterized in terms of the two kinds of non-local order
parameters, e.g., \textit{the expectation value of (open or close)
string-like operators }$\prod\limits_{i_n=1}^i\sigma _{i_n}^x$ and $%
\prod\limits_{_{i_n=1}}^iF_{i_n}$. And the topological ordered state are
destroyed by the condensation of the open-string operators ($Z_2$ charge and
$Z_2$ vortex). These results indicate a new type of QPT between a
topological ordered phase and a non-topological ordered phase.

The authors acknowledge stimulating discussions with S. Chen, N.H. Tong.
S.P. Kou acknowledges that this research is supported by NFSC Grant no.
10574014.


\begin{thebibliography}{99}
\bibitem{commet}  A small \textrm{y}-direction transverse field leads to a
non-local-interchain interaction for the dual $d=1$ Ising chains. Such
relevant interaction couples isolated Ising chains together. And the effects
of it will be explored elsewhere.
\end{thebibliography}
\end{document}